\documentclass[twocolumn,a4paper,superscriptaddress,floatfix]{revtex4}

\usepackage{amsmath,color,epsfig,setspace}

\begin{document}

\title{Covalent bonds against magnetism in transition metal compounds.}

\author{Sergey V.~Streltsov}
\affiliation{Institute of Metal Physics, S. Kovalevskoy St. 18, 620990 Yekaterinburg, Russia}
\affiliation{Department of theoretical physics and applied mathematics, Ural Federal University, Mira St. 19, 620002 Yekaterinburg, Russia}

\author{Daniel I.~Khomskii}
\affiliation{II. Physikalisches Institut, Universit\"at zu K\"oln, Z\" ulpicher Stra\ss e 77, D-50937 K$\ddot o$ln, Germany}


\def\red{\color{red}}
\def\blue{\color{blue}} 
\def\green{\color{green}} 

\begin{abstract}
Magnetism in transition metal compounds is usually considered starting
from a description of isolated ions, as exact as possible, and treating their
(exchange) interaction at a later stage. We show that this standard
approach may break down in many cases, especially in $4d$ and $5d$
compounds. We argue that there is an important intersite effect -- an
orbital-selective formation  of covalent metal-metal bonds, which
leads to an ``exclusion'' of corresponding electrons  from the magnetic
subsystem, and thus strongly affects magnetic properties of the
system. This effect is especially prominent for noninteger electron
number, when it results in suppression of the famous double exchange,
the main mechanism of ferromagnetism in transition metal compounds. We
study this novel mechanism analytically and numerically and show that
it explains magnetic properties of not only several $4d-5d$ materials,
including Nb$_2$O$_2$F$_3$ and
Ba$_5$AlIr$_2$O$_{11}$, but can also be operative in $3d$
transition metal oxides, e.g. in CrO$_2$ under pressure. We also discuss the role of spin-orbit coupling on the competition between covalency and magnetism. Our results
demonstrate that strong intersite coupling may invalidate the
standard single-site starting point for considering magnetism, and can
lead to a qualitatively new behaviour. 
\end{abstract}

\maketitle

\section{Introduction}
Transition metal (TM) compounds present one of the main playgrounds in a large field of magnetism\cite{Imada1998,khomskii2014transition,sachdev}. Usually, when considering magnetic properties of these systems, one starts from the, as exact as possible, treatment of isolated TM ions or such ions in the surrounding of ligands, e.g. TMO$_6$ octahedra. A typical situation for a moderately strong crystal field is the one in which $d-$electrons obey the Hund's rule, forming a state with the maximal spin. For a stronger crystal field low-spin states are also possible, but also in this case electrons in degenerate subshells, e.g. $t_{2g}$ electrons, first of all form a state with maximal possible spin. Then, these large total spins interact by exchange coupling with the neighbouring TM ions. This interaction, a superexchange for integer electron occupation\cite{Anderson1959}, or a double exchange  for partially-filled $d$-levels\cite{Zener1951}, is then treated  using this starting point with this total spin of isolated ions, taking into account the hopping between sites (leading  in effect  to magnetic interaction) as a weak perturbation, which does not break the magnetic state of an ion.

For heavier elements, such as $4d$ or $5d$ TM, one should also take into account the relativistic spin-orbit (SO) coupling, which couples the total spin $S$ as dictated by the Hund's rule, with the (effective) orbital moment $L$. But, in any case, it is usually assumed that the ``building blocks'' for further consideration of the magnetic interactions are such isolated TM  ions with the corresponding quantum numbers.

However, especially when we go to heavier TM ions, such as $4d$ and $5d$, also the spatial extent of the corresponding $d-$orbitals increases strongly, and with it the effective $d-d$ hopping, $t$
\cite{khomskii2014transition}. One can anticipate a possibility for this hopping to become comparable with or even exceed the intra-atomic interactions, such as the Hund's coupling, $J_H$, and spin-orbit coupling $\lambda$. In such a case the standard approach described above may break down and one has to include intersite effects from the very beginning. We claim that this indeed  happens in many $4d$ and $5d$ systems with appropriate geometries. The resulting effect is that, for example, in a TM dimer with several $d-$electrons per ion, some electrons, namely those occupying the orbitals with the strongest overlap, form intersite covalent bonds, i.e. the singlet molecular orbitals (MO). In this case such electrons become essentially nonmagnetic and ``drop out of the game'', so that only $d-$electrons in orbitals without such a strong overlap may be treated as localized, contributing to localized moments and to eventual magnetic ordering. The result would be that the  effective magnetic moment  of a TM ion in such situation would become much smaller than the nominal moment corresponding to the formal valence of an ion. Below we demonstrate that this effect is indeed  realized in many TM compounds, especially those of $4d$ and $5d$ elements. And, besides reducing the magnetic moment of an ion, this mechanism can suppress the notorious double exchange (DE) mechanism of (ferro)magnetic ordering $-$ the main mechanism of ferromagnetism in systems with a fractional occupation of the $d-$levels.

Before presenting main results let us emphasize that the interplay between the formation of covalent bonds, spin-orbit coupling and intra-atomic exchange interaction discussed in the present paper is very important for intensively studied nowadays $4d$ and $5d$ transition metal oxides such as $\alpha-$RuCl$_3$~\cite{Plumb2014,Sears2015}, Li$_2$RuO$_3$\cite{Kimber2013,Wang2014}, LiZn$_2$Mo$_3$O$_8$~\cite{sheckelton2012,Mourigal2014} and Na$_2$IrO$_3$\cite{Chaloupka2010,Choi2012,Foyevtsova2013}. Competition of these interactions results in highly unusual physical properties in these systems: Kitaev spin liquid, valence-bond condensation and different topological effects.

\section{Qualitative considerations.}
\begin{figure}
\begin{center}
\includegraphics[angle=0,width=1\columnwidth]{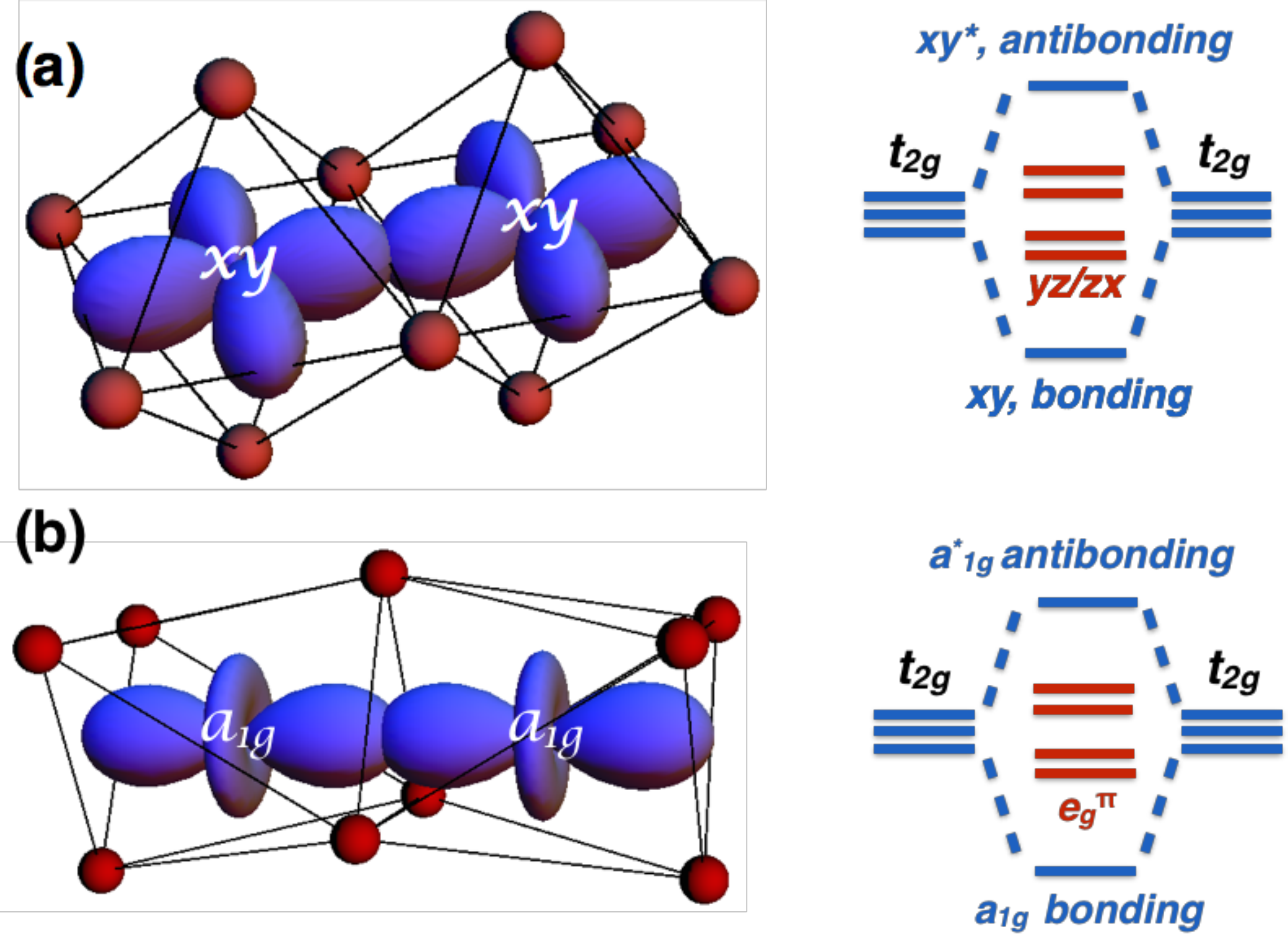}
\end{center}
\caption{\label{ORB}  Orbitals having large overlap ($c-$orbitals in our notations) and corresponding level splitting with the formation of bonding and antibonding orbitals for transition metal dimers
in different geometries: (a) common edge geometry and $xy-$orbitals, (b) common face geometry  and $a_{1g}-$orbitals.
}
\end{figure}

We start by simple qualitative arguments, considering a two site problem with, for example, three electrons per dimer, occupying two types of orbitals: orbital $c$ (the corresponding creation and annihilation electron operators at the site $i$ with the spin $\sigma$ are $c^{\dagger}_{i \sigma}$, $c_{i \sigma}$), with intersite hopping $t_c$, and another orbital, $d$, with a very weak hopping $t_d$ (which we for simplicity at the beginning put to zero). In reality the $c-$orbitals could be, for example, the strongly overlapping $xy$ orbitals in the situation when the neighbouring TMO$_6$ octahedra share  edges, i.e. have two common oxygens, see Fig. \ref{ORB}(a), or the $a_{1g}$ orbitals in the common face geometry, Fig. \ref{ORB}(b); the orthogonal orbitals ($zx$, $yz$ in the first case or $e_g^{\pi}$ orbitals in the second) would then play a role of localized $d-$orbitals (for a more detailed treatment of these situations see, e.g., in Refs.~\cite{Goodenough, Kugel2015, Khomskii2016}).

\begin{figure}[t!]
 \begin{center}
\includegraphics[angle=0,width=1\columnwidth]{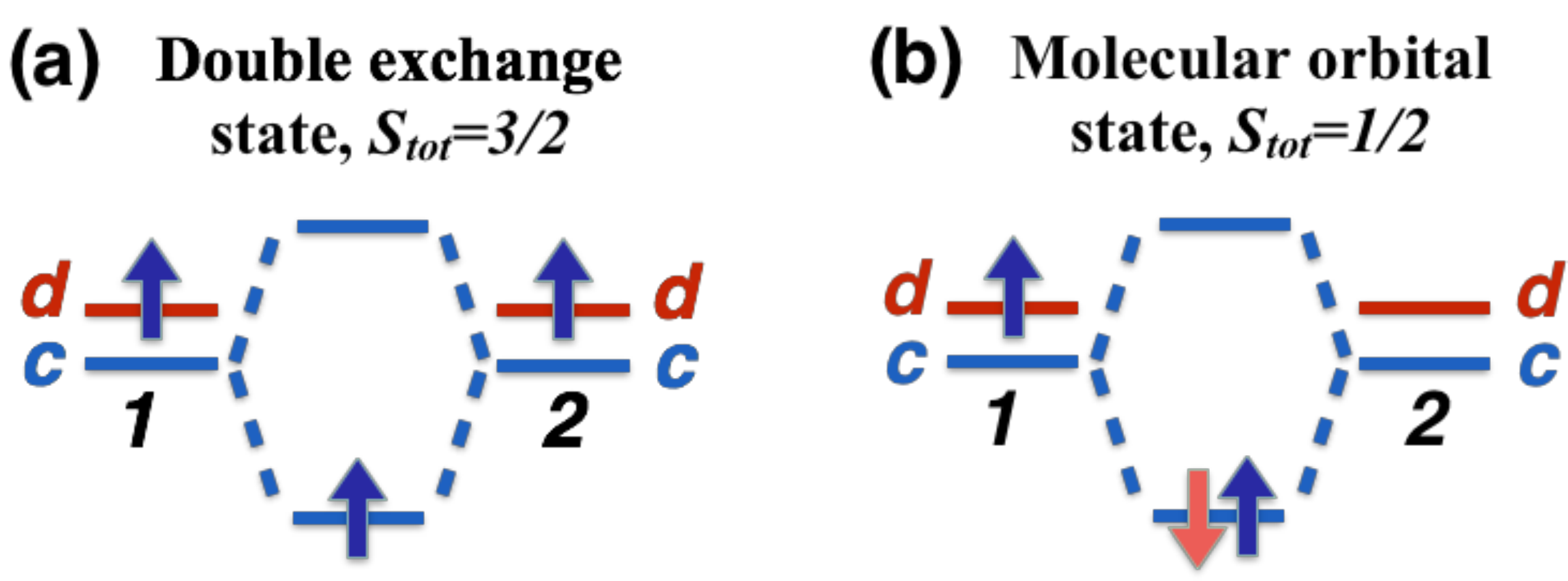}
\end{center}
\caption{\label{Level-schema} 
Two competing electronic configurations with different total spin $S_{tot}$: 
(a) DE and (b) MO states for a dimer with 2 orbitals ($c$ and $d$) per site 
and 3 electrons. 
}
\end{figure}

In this case we can consider two different limiting states. In the state depicted in Fig.~\ref{Level-schema}(a) 
we put two electrons into the localized $d-$orbitals, and the remaining electron occupies the ``itinerant'' $c-$orbital, hopping back and forth from site 1 to site 2, or occupying the bonding state $ ( c^{\dagger}_{1 \uparrow} +c^{\dagger}_{2 \uparrow})/\sqrt 2$. To optimize the Hund's intra-atomic exchange, this ``hopping'' electron would have its  spin parallel to the spins of the localized electrons, and its delocalization stabilises the state with maximum spin, here $S_{tot}=3/2$. This is, in essence, the double exchange mechanism of ferromagnetism, first proposed by Zener
for just such a dimer\cite{Zener1951} .
It is easy to see that the energy of this DE state is 
\begin{eqnarray}
\label{DE-energy}
E_{DE} = -t_c - J_H.                           
\end{eqnarray}
For simplicity we ignore here the contribution of the on-site Coulomb (Hubbard) interaction, see a more complete treatment below. We also treat the Hund's interaction 
\begin{eqnarray}
\label{Hund}
H_{Hund} = - J_H (1/2 + 2 \vec S_1 \vec S_2)
\end{eqnarray}
in the mean-field approximation, where $ \vec S_1$ and $ \vec S_2$ are the total spins on sites
1 and 2.

This DE state is definitely the most favourable state for a strong Hund's interaction. And, indeed, in all papers on double exchange \cite{Zener1951,DeGennes1960,Anderson1955} 
one makes such an approximation, most often by simply setting $J_H$ to infinity. 
This is a reasonable assumption for  most $3d$ compounds at ambient conditions, for which 
$J_H \sim $0.7-0.9 eV, usually (much) larger than the effective $d-d$ hopping (direct, or via ligands), which is typically 0.1-0.3 eV. However, when we go to $4d$, and especially $5d$ systems, the situation may change drastically: $J_H$ is reduced \cite{VanderMarel1988} (to $\sim $0.5-0.6 eV for $4d$ and $\sim$0.5 eV for $5d$ elements\cite{Sasioglu2011}),
 whereas the spatial extension of $d-$wave function, and the corresponding overlap and $d-d$ hopping strongly increase, so that $t_c$ can easily be of order of 1.0-1.5 eV, as e.g. in Li$_2$RuO$_3$\cite{Kimber2013} or Y$_5$Mo$_2$O$_{12}$\cite{Streltsov2015MISM}. In this case we can form a different state (illustrated in Fig.\ref{Level-schema}(b)), redistributing electrons between $d-$orbitals: we now put two electrons on the ``itinerant''  $c-$orbitals, so that they form a singlet state in the bonding orbital of Fig.~\ref{Level-schema}(b) (actually a molecular orbital state):
\begin{eqnarray}
\label{MO-WF}
| MO \rangle = ( c^{\dagger}_{1 \uparrow} +c^{\dagger}_{2 \uparrow})( c^{\dagger}_{1 \downarrow} +c^{\dagger}_{2 \downarrow})/2
\end{eqnarray}
leaving one electron on localized $d-$orbitals. In effect, the  total spin of this state (we call it for short the MO state) is not 3/2, as for the DE state of Fig.~\ref{Level-schema}(a), but only $S_{tot}=1/2$! The energy of this state in our simple approximation is
\begin{eqnarray}
\label{MO-energy}
E_{MO} = -2t_c - J_H/2.                           
\end{eqnarray}
Comparing the energies (\ref{DE-energy}) and (\ref{MO-energy}), we see that the MO state with suppressed double exchange and strongly reduced total spin is more favourable if
\begin{eqnarray}
\label{Jc-naive}
2t_c > J_H.          
\end{eqnarray}
Thus, we observe that in this simplified model the covalent bonding, defined by the hopping $t_c$ (which can include also the hopping via ligands), can suppress the DE.

One can use a more general wave function of MO type, with two electrons in the $c-$orbitals  with the total $S=0$, and one localized electron with $S=1/2$ in $d-$orbitals. For example, if we put $d-$electron with spin $\uparrow$ at a site 1, we can take generalized MO state in the form
\begin{eqnarray}
| \widetilde {MO} \rangle = (\alpha c^{\dagger}_{1 \uparrow} + \beta c^{\dagger}_{2 \uparrow})( \beta c^{\dagger}_{1 \downarrow} + \alpha c^{\dagger}_{2 \downarrow})/2                                      
\end{eqnarray}
with the variational parameters $\alpha$, $\beta$ such that $\alpha > \beta$, so as to win the Hund's exchange energy with the $d_{1\uparrow}$ electron (at the expense of some loss of the bonding energy). This would stabilize the MO (or rather $\widetilde {MO}$) state ever more, shifting the critical value $J_H^c$ needed for stabilizing DE state to larger values (i.e.  $J_H^c \sim 6t_c$, and not $J_H^c \sim 2t_c$ as follows from Eq. (\ref{MO-WF})). I.e., the electron hopping can be even more efficient in suppressing DE than it follows from the simple estimate in Eq. (\ref{Jc-naive}).
\begin{figure}[t!]
 \begin{center}
\includegraphics[angle=0,width=1\columnwidth]{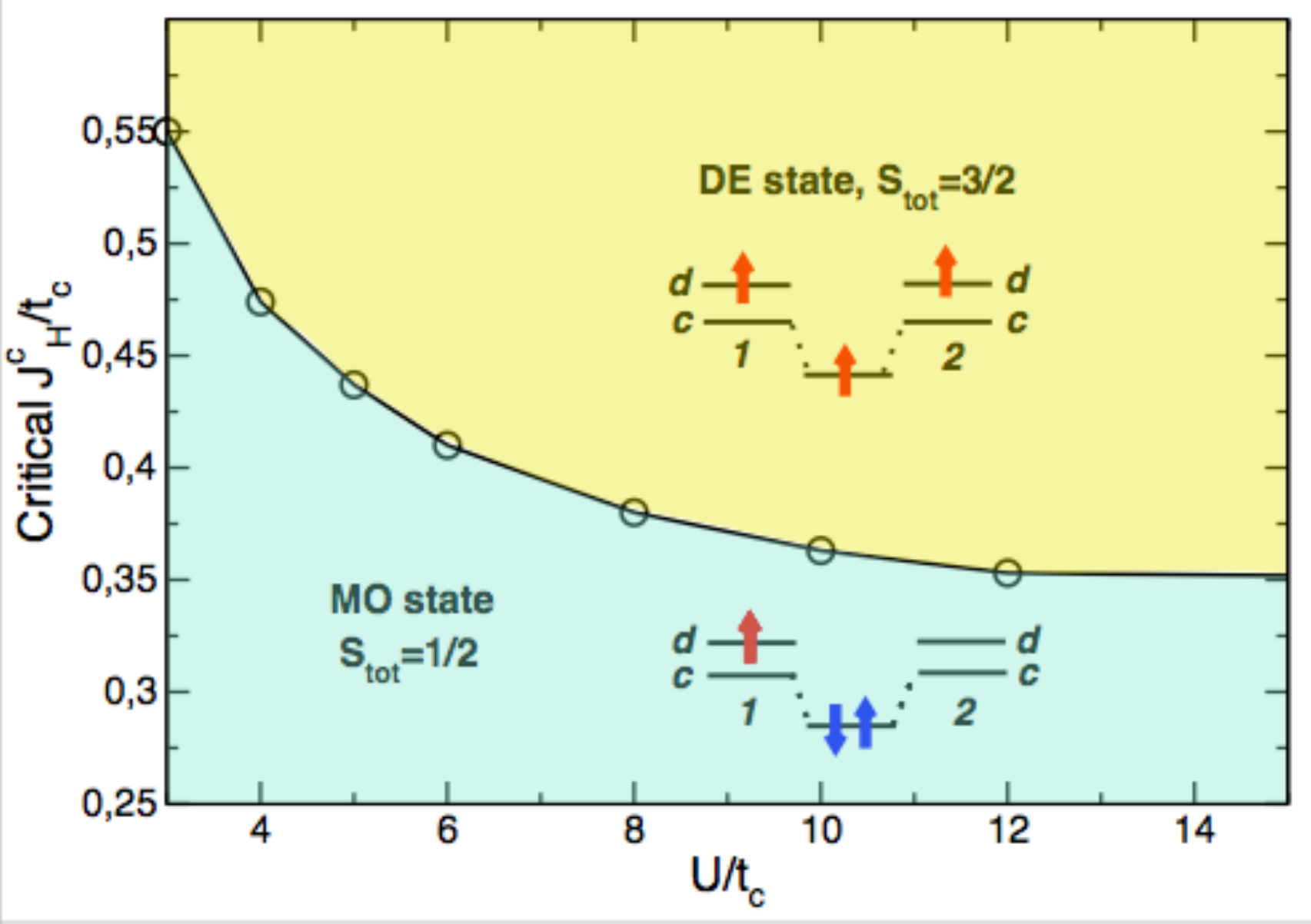}
\end{center}
\caption{\label{PD} Phase diagram for a dimer with 2 orbitals ($c$ and $d$) per site and 3 electrons in $U$ (on-site Hubbard repulsion) and $J_H$ (Hund's rule coupling) coordinates; $c$-orbitals  form molecular-orbitals (corresponding hopping is $t_c$), while $d$-orbitals stay localized ($t_d=0.1 t_c$, the same as in Fig. \ref{S2}). Results of the exact diagonalization at $T=0$ K. We restrict ourself only by the
repulsive Coulomb potential, which in the Kanamori parametrization corresponds to $U > 3J_H$. \bigskip
}
\end{figure}

\section{Model consideration}
 In fact such a model with two orbitals per site and three electrons per dimer 
can be solved exactly (see the details in Supporting information; here and below we used the Hubbard model within the Kanamori parametrization rather than a simplified expression for the interaction term given in \eqref{Hund}), including also the on-site Coulomb (Hubbard) repulsion $U$, besides the
Hund's rule exchange $J_H$ and the hopping of $d-$ and $c-$orbitals ($t_c$ and $t_d$).
The corresponding phase diagram for $T=0$ K is shown in Fig. \ref{PD}. One may see that indeed the MO state can be realized 
for any $U$, if $J_H$ is small enough.  This state can be considered as orbital-selective\cite{Streltsov2014}, since only $d-$orbitals provide local moments, while electrons on $c-$orbitals form spin-singlet. For a finite $U$ the simple estimates like Eq.~(\ref{Jc-naive}) do not hold anymore, and the critical value $J_H^c$ needed to stabilise the DE state is much smaller than for $U=0$,
since the Coulomb repulsion modifies the ground-state wave function for $c-$electrons
from real molecular-orbital to the Heitler-London, or rather Coulson-Fisher-like, with an increased weight of the ionic terms (with respect to homopolar ones)\cite{Coulson1949}. The transition from one state to another is discontinues, since they are characterised by different quantum numbers  ($S_{tot}=3/2$ for DE state vs. $S_{tot}=1/2$ for MO state), 
and corresponding terms do not hybridize, but simply cross.

\begin{figure}[t!]
 \begin{center}
\includegraphics[angle=0,width=0.98\columnwidth]{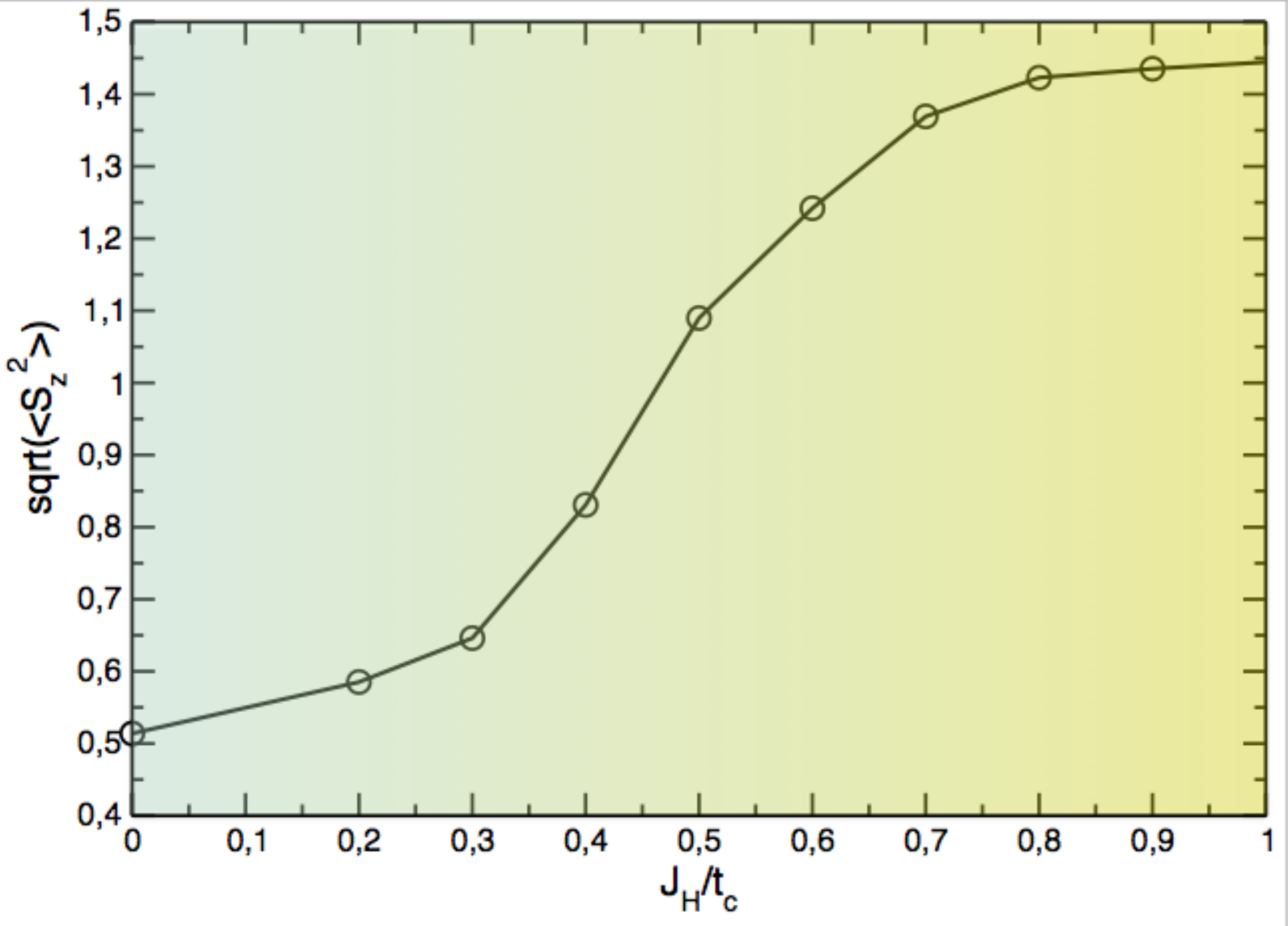}
\end{center}
\caption{\label{S2} Dependence of $\sqrt {\langle S^2_{tot,z} \rangle}$ on the $J_H/t_c$ ratio at fixed
$U=6 t_c$ for a dimerized chain with 1.5 electrons and with 2 orbitals per site at $T=1100$ K. Intra-dimer hopping parameter for $d-$orbitals was taken to be $0.1 t_c$, while inter-dimer hopping $t'_c = t'_d = 0.05 t_c$. Results of the cluster DMFT calculations.}
\end{figure}

We can generalize this treatment onto larger systems, 
which we did for a dimerized chain with two orbitals and 1.5 electrons per site. 
The calculations have been performed
using a cluster version of the dynamical mean-field theory (DMFT)\cite{Kotliar2001}. The cluster DMFT was shown to provide a very accurate description of the one-dimensional
chain \cite{Kotliar2001}. The results of the cluster DMFT calculations for a dimerized chain at a fixed $U$ are shown in Fig. \ref{S2}. We see that indeed the MO-like state with $S_{tot}=1/2$ survives for realistic values of $J_H$ (for $4d-5d$ materials)\cite{VanderMarel1988,Sasioglu2011} and the DE can be suppressed for small $J_H$ even in an extended system. The transition to this MO state is not discontinuous now, but is broadened, because of two factors: temperature (which is one of the parameter in our cluster-DMFT calculations)
and formation of bands, or in other words the presence of the finite electron hoppings between dimers. As it was explained above, the critical Hund's rule coupling strongly depends on $U$. If $U$ is small, then $J_H$ competes with the difference of hopping parameters, $t_c - t_d$, which defines the energy of the MO. In contrast, with increasing $U$ we quickly arrive at the Heitler-London type of the wave function for $c-$electrons with the energy gain $\sim t^2_c/U$ in the MO state. 

\section{Role of the spin-orbit coupling \label{sec:soc}}
As the effects discussed in this paper are met mainly in $4d$ and $5d$ compounds (although not exclusively, see the discussion about CrO$_2$ in Sec.~\ref{sec:real-mat}), it is important to address the possible role of spin-orbit coupling (SOC), which is strong in these systems. It turns out that the effect of SOC is not universal and depends on a specific situation.
 \begin{figure}[t!]
 \begin{center}
\includegraphics[angle=270,width=1\columnwidth]{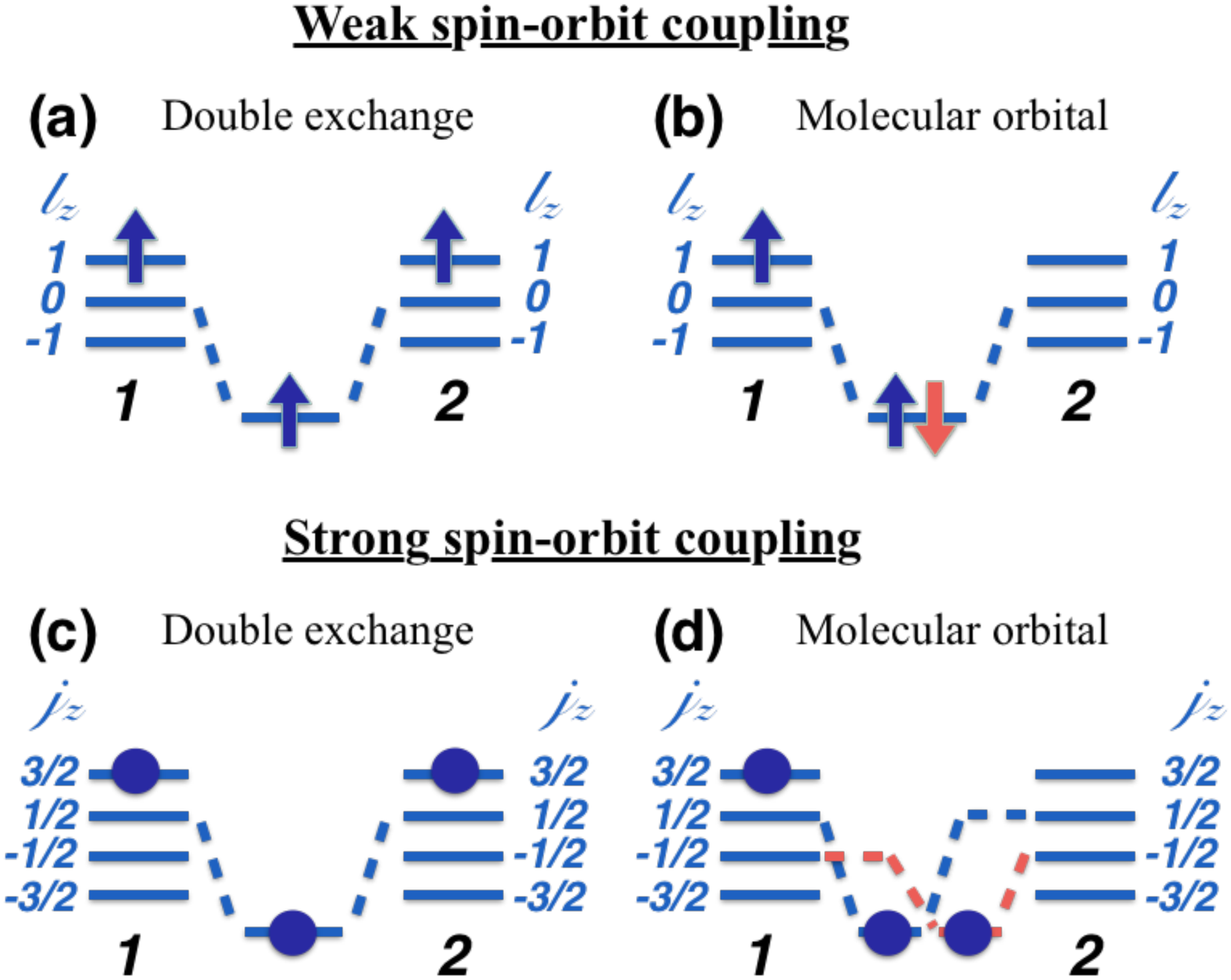}
\end{center}
\caption{\label{Level-schema-SOC} 
Two competing electronic configurations for a dimer- DE and MO in the case of weak and strong spin-orbit coupling, three $t_{2g}$ orbitals  per site and three electrons. 
}
\end{figure}

Consider first the same situation with three electrons per dimer and three-fold degenerate $t_{2g}$ 
orbitals, which now can be labeled by value of $z-$projection of the effective orbital moment $\vec l = 1$, i.e. $|l^z=0 \rangle \equiv |0 \rangle$ and $|l^z = \pm 1 \rangle \equiv |\pm 1 \rangle$.  For the sake of simplicity here and below we consider SOC on the one-electron level, i.e. $H_{SOC}=-\sum_m \zeta \vec l_m \vec s_m$ (here $m$ numerates orbitals). Then for weak SOC, $\zeta \ll (t_c, J_H)$ , we can again have two situations: that of the DE, with the maximum spin for a dimer possible, here $S_{tot}=3/2$, Fig.~\ref{Level-schema-SOC}(a), and the state with the singlet MO and the total spin $S_{tot}=1/2$, Fig. ~\ref{Level-schema-SOC}(b). To gain at least some SO energy, we put localized electrons with spin $\uparrow$ on the orbital $|+1 \rangle$. Since in the most geometries such as edge and face-sharing the orbitals with $l_z=0$ ($xy$ and $a_{1g}$ orbitals respectively) strongly overlap, we put  ``itinerant'' electrons on the bonding orbital $\frac 1 {\sqrt 2} (|0 \rangle_1 + |0 \rangle_2)$, where 1 and 2 are site indexes.  The energy of the DE state in this case is
\begin{eqnarray}
\label{DE-energy-SOC}
E^{DE}_{weak-SOC} = -t_c - J_H - \zeta, 
\end{eqnarray}
(cf.~\eqref{DE-energy}), 
since here for localized electrons only the``classical'' part of the SOC contributes in lowest order, -$\zeta \langle 1,\uparrow| l^zs^z | 1,\uparrow \rangle=-\zeta/2$ per site, while in the lowest order the bonding states with $l_z=0$ do not give energy gain due to SOC (it will appear due to ``quantum'' terms $l^+s^-$ etc. in the second order in $\zeta/t_c$). Similarly, the energy of the MO state of Fig.~\ref{Level-schema-SOC}(b) is 
\begin{eqnarray}
\label{MO-energy-SOC}
E^{MO}_{weak-SOC} = -2t_c - J_H/2 - \zeta/2, 
\end{eqnarray}
cf.~\eqref{MO-energy}. Comparison of these expressions shows that in this case SOC stabilizes magnetic DE state: the condition for low-spin MO state is now, instead of \eqref{Jc-naive},
\begin{eqnarray}
\label{Jc-naive-SOC}
2t_c > J_H + \zeta.                                                                                                                               
\end{eqnarray}
This agrees with results of Ref.~\cite{Isobe2014}, where the case of the weak SOC was considered.

However, the situation is very different for strong SOC,  $\zeta \gg (t_c, J_H)$. In this case we can project relevant states onto a quartet $j=\frac  3 2$, with the states 
$|j^z= \frac 3 2 \rangle = |1, \uparrow \rangle$,  
$|j^z=\frac  1 2\rangle = \sqrt {\frac 2 3} |0, \uparrow \rangle + \sqrt{\frac 1 3}|1, \downarrow \rangle$,  
$|j^z=- \frac 1 2 \rangle = \sqrt { \frac 2 3} |0, \downarrow \rangle + \sqrt{\frac 1 3} |-1,\uparrow \rangle$,  $|j^z=- \frac 3 2 \rangle = |-1, \downarrow \rangle$.  
Then the “DE” state with the maximum total moment $j$ of a dimer 7/2 is the state shown in Fig.~\ref{Level-schema-SOC}(c), with localized $d-$electrons on sites 1 and 2 in states $|j^z=3/2 \rangle$, and a  ``mobile'' electron - on a bonding orbital $\frac 1 {\sqrt 2} (|j^z=1/2\rangle_1 + |j^z=1/2\rangle_2)$. For such state the effective hopping is reduced, as only the $|0 \rangle$ component ``hops'' to a neighbour, $\langle j^z=  \pm \frac  1 2 | \hat t | j^z= \pm \frac  1 2 \rangle = \frac 2 3  \langle 0| \hat t |0 \rangle = \frac 2 3 t_c$, and the total energy of such state turns out to be  
\begin{eqnarray}
E^{DE}_{strong-SOC} = -\frac 2 3 t_c - \frac 3 2 \zeta - \frac 2 3  J_H                                                                      
\end{eqnarray}
(SOC energy in any state of the quartet $j=\frac 3 2$ is  $-\zeta/2$ per electron; the calculation of the Hund's energy contribution is quite straightforward, but somewhat tedious \footnote{We treat the Hund’s energy in a mean-field approximation, using the ``recipe'' described for example in Ref.~\cite{khomskii2014transition}: the Hund's energy is equal to
 $J_H \times$(number of parallel spins). More accurate treatment of the Hund’s coupling, including quantum effects in it, would somewhat change numerical coefficients, but would not change qualitative conclusions. In our numerical and {\it ab initio} calculations presented below the Hund’s rule coupling is taken into account in a full spherically-symmetric form, including quantum effects in it.}).

Similarly, an analogue of the MO state is shown in Fig.~\ref{Level-schema-SOC}(d). In this state we put one electron on a localized state $|j^z=\frac 3 2 \rangle$, say on a site 1, and two electrons on bonding orbitals $\frac 1 {\sqrt 2}(|j^z=\frac 1 2\rangle_1 + |j^z=\frac 1 2 \rangle_2)$ and $\frac 1 {\sqrt 2} (|j^z=-\frac 1 2\rangle_1 + |j^z=-\frac 1 2\rangle_2)$. The energy of this state, calculated similarly to that of the DE state in this regime, is
\begin{eqnarray}
E^{MO}_{strong-SOC} = - \frac 4 3 t_c - \frac 3 2 \zeta - \frac {13} {18} J_H.                                                                    
\end{eqnarray}
We see that in this regime (strong SOC) the MO state is definitely favoured over the DE state: the contribution from the SOC in any state of a quartet is the same, and in the MO state both the bonding  and even Hund’s energies are lower.

Summing up, in the case of 3 electrons per dimer weak SOC acts in favour of the magnetic DE state, but strong SOC, instead, stabilizes the MO state with reduced moment.
This is related to the fact that the energy of the bonding orbital (the
lowest curve in Fig. S5) rapidly decreases with the increase of SOC, which
makes the MO state (in which this orbital is occupied twice) more favourable.

\begin{figure}[t!]
 \begin{center}
\includegraphics[angle=0,width=0.85\columnwidth]{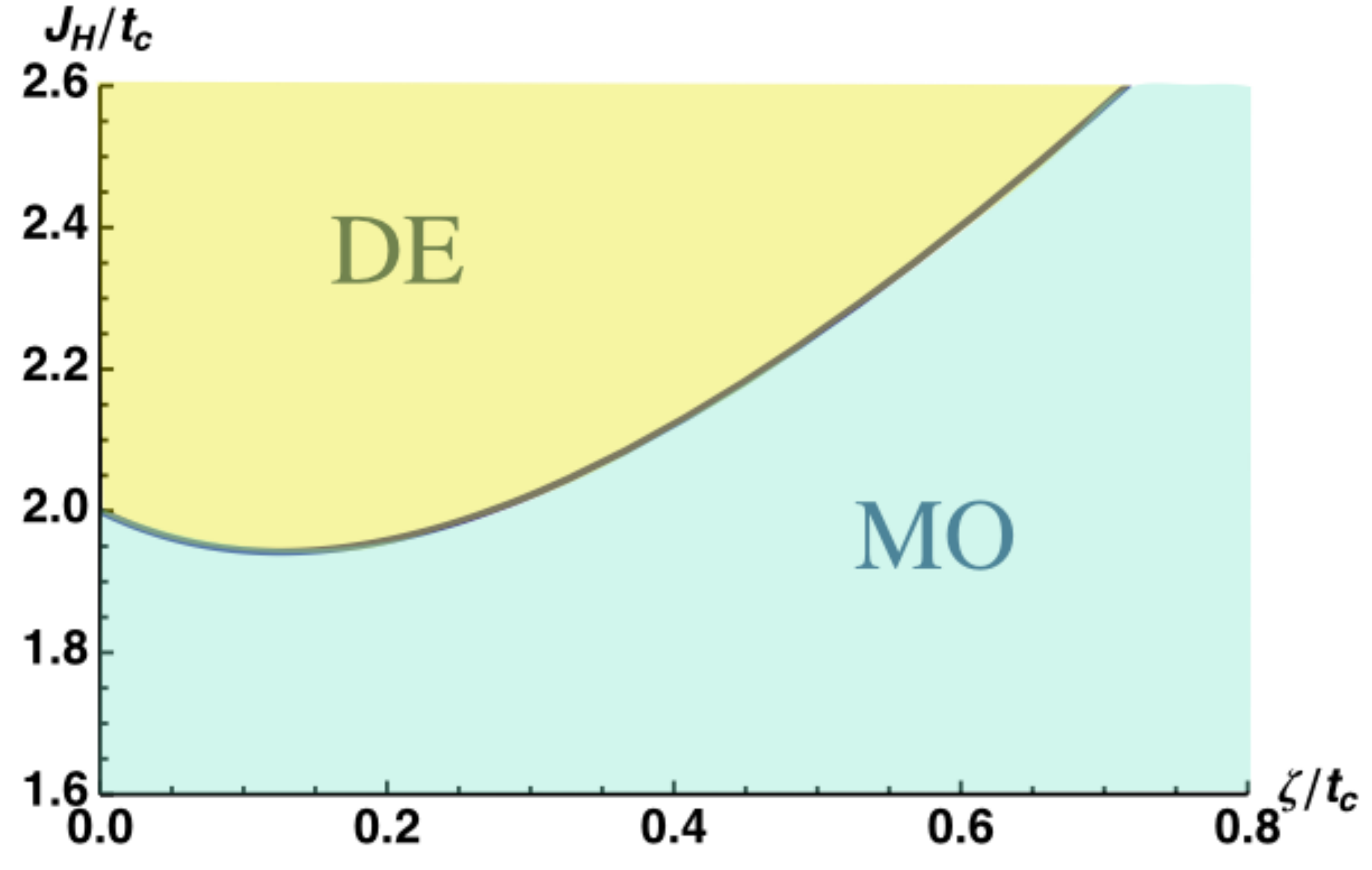}
\end{center}
\caption{\label{PD-SOC-3} Phase diagram for a dimer with 3 $t_{2g}$ orbitals per site and 3 electrons, which takes into account spin-orbit coupling on the one-electron level. Results of the exact diagonalization at $T=0$ K. }
\end{figure}

The analytic treatment presented above is confirmed by the exact diagonalization results for the dimer with 3 orbitals per site. The resulting phase diagram is shown in Fig. \ref{PD-SOC-3}. We see that it agrees with our analytical results for limiting cases of $\zeta \to 0$ and $\zeta \to \infty$ presented above, and it has a ``reentrant'' character: for certain values of parameters the increase of SOC can initially transform system into the magnetic DE state, but for larger $\zeta$ - back to the MO state.

One can show that for the weak SOC the situation is the same also not for three electrons, but for three holes in $t_{2g}$ levels (situation typical, for example, for a dimer of, formally, Ir$^{4+}$($d^5$) and Ir$^{5+}$ ($d^4$)): the weak SOC works in favour of a “more magnetic” DE state. However, for strong SOC we do not have anymore the electron-hole symmetry: as is well known \cite{khomskii2014transition,Abragam}, these states can be projected onto a Kramers doublet $j=1/2$ without any extra orbital degeneracy, and for the situation with $d^4/d^5$ occupation (9 electrons per dimer) only the analogue of a low-spin state with the total moment 1/2 is possible. Thus, in this case the qualitative phase diagram would again have the form similar to the one shown in Fig.~\ref{PD-SOC-3}, with weak SOC stabilising magnetic DE state, but strong SOC suppressing total moment - although mathematical description is different from the case with three $d-$electrons.

However, for example one can show that the situation is qualitatively different for 5 electrons per dimer: in this case the treatment similar to that for three electrons above, shows that both strong and weak SOC works in favour of a less-magnetic MO state, i.e. the corresponding curve separating MO and DE states in the phase diagram, similar to that of Fig.~\ref{PD-SOC-3}, would increase already at small $\zeta$, without reentrant behaviour. Thus, we see that, in contrast to the Hund’s coupling, which always competes with the electron hopping and tends to stabilize a ``more magnetic'' state, the SOC can work differently in different situations. 

One more important factor, which can change the interplay of SOC with the Hund’s exchange and electron hopping, is a possible contribution of not only the direct $d-d$ hopping, considered in our treatment, but also of the hopping via  $p-$orbitals of ligands, e.g. oxygens. To take into account all these effects, for real materials it is probably better to rely on {\it ab initio} calculations, the results of some of which are presented in the next section.

\section{Suppression of the double exchange and magnetic moment in real materials.\label{sec:real-mat}} 

We now turn to real materials and show that the physics described above (strong reduction of magnetic moment and eventual suppression of double exchange, due to formation of orbital-selective  covalent bonds between TM) indeed works in real materials and allows us to explain the behaviour of many $4d$ and $5d$ systems. The first example of such a system
is the recently synthesized oxyfluoride Nb$_2$O$_2$F$_3$ \cite{Tran2015}
(see its structure and the results of our {\it ab initio} calculations in Supporting information). While, according to the chemical formula there has to be  three $d-$electrons per Nb dimer, i.e. $S_{tot} = 3/2$, the experimental effective magnetic moment is consistent with $S_{tot}=1/2$ per dimer \cite{Tran2015}. 
The band-structure calculations show that the bonding-antibonding splitting 
for the $xy$ orbitals pointing to each other in the common edge geometry (cf. Fig. \ref{ORB}(a)) is 
very large here, $2t_c \sim 3$ eV -- 
much larger than the Hund's rule energy estimated to be $J_H \sim $0.5 eV for Nb\cite{Sasioglu2011}.

According to the results of our model cluster DMFT calculations, see above,  for $J_H/t_c \sim 0.3$ the DE is largely suppressed, which agrees with the experimental findings \cite{Tran2015}. However, parameters $t_d$, $t'$, $U$ in Nb$_2$O$_2$F$_3$ can be very different from what we used in the model
calculations, so that comparison with the results of the GGA (generalized gradient approximation) should be more appropriate. These results are summed up in Supplementary materials and
indeed show that the value of the local magnetic moment in the high temperature phase is $\mu^{theor} \sim 1 \mu_B$ per dimer, which agrees with the experimental value of the effective moment \cite{Tran2015,Gapontsev2016}.

We have discussed so far how covalent bonding competes with Hund's rule coupling. However, for $4d$ and $5d$ TM compounds 
also another ``player'' can enter the game: the spin-orbit coupling, which is known to be strong enough in heavy $4d/5d$ metals and has been already demonstrated to cause very unusual physical effects \cite{Kim2008,Jackeli2009,Witczak-Krempa2014}. 
On our third example, Ba$_5$AlIr$_2$O$_{11}$, we show that this interaction may indeed
take part in the competition between intra-atomic exchange and covalent bonding.

The crystal structure of Ba$_5$AlIr$_2$O$_{11}$ consists of [Ir-Ir]$^{9+}$ dimers (Fig. S4(a) in Supplementary materials) with
9 electrons or 3 holes per dimer, so that this is exactly the example of a system,
considered in the first part of our paper,
 where the DE and MO states could compete. And, indeed, the effective magnetic moment in the Curie-Weiss law $\mu_{eff} \sim 1 \mu_B$/dimer,
measured at high temperatures, indicates substantial suppression of the magnetic 
moment~\cite{Terzic2015}. However, our GGA calculations show that accounting for covalency and exchange splitting is not sufficient to explain this small $\mu_{eff}$; they  give $\mu^{GGA}\sim 2 \mu_B$. 
It is the spin-orbit coupling, that, by additional splitting of $d$ levels, 
conspires with the covalency and finally helps to suppress the DE in this system, see Supplementary material. However, as stressed in Sec.~\ref{sec:soc}, this is not a general conclusion: depending on the filling of the $d-$shell and on a specific geometry, the spin-orbit interaction can act against the DE or support it.

Until now we have considered real materials in which the structure provides relatively well separated dimers. But, as we saw, e.g., in our cluster DMFT calculations, the effect of suppression of the DE can survive even for solids, without well-defined dimers. In some of such systems singlet covalent bonds can form spontaneously, also acting against DE (cf. Ref. \cite{Nishimoto2012}).

One such example seems to be given by CrO$_2$ $-$ a classical DE system \cite{Korotin1998}. In CrO$_2$ one of the two $3d-$electrons is localized in the $xy$ orbital on each Cr site, while another one occupies a broad $xz/yz$ band, stays itinerant and actually makes this system
ferromagnetic\cite{Korotin1998}. In contrast to VO$_2$, having the same rutile structure, 
but one $d-$electron per V, CrO$_2$ does not dimerize at zero pressure. 

Apparently, in normal conditions in this system the hopping via oxygens is more important, and in effect DE wins and provides ferromagnetism with large Curie temperature. However, under high pressure the situation can change: the direct $d-d$ hopping can begin to dominate, the bonding-antibonding splitting even in this 3D system can become comparable with what we have in $4d$ and $5d$ TM compounds at normal conditions,
in which case our physics could come into play. And indeed,
accurate band structure calculations\cite{Kim2012a} have found a cascade of structural transitions in CrO$_2$ with pressure, with the formation of the dimerized monoclinic structure for P$\sim$ 70 GPa . 
As in the case of the Jahn-Teller distortions, it is not clear whether the lattice or electronic subsystem triggers the structural transition, but once the dimerization starts they go hand in hand and destabilize the DE state by lowering the energy of the $xy$ molecular orbital. At some point two out of four $d-$electrons in the Cr-Cr dimer occupy this molecular orbital, while the remaining two electrons provide metallicity and eventual paramagnetism,
similar to what is observed in MoO$_2$ at normal conditions \cite{Eyert2000} or expected in MoCl$_4$ at moderate pressure\cite{Korotin2016}.
We thus see that even some $3d$ systems could in principle be tuned to the MO regime, in particular under pressure, which increases $d-d$ overlap and hopping, making such systems more similar to the $4d$ and $5d-$ones. 

It is also possible that some other factors, such as charge ordering\cite{Gapontsev2016}, could modify the situation. 
The external perturbations, such as temperature or pressure, can drive a system from the DE to the MO state, so that the phase diagrams  of  systems with competing DE and MO states can be rather rich and their physical properties can be highly unusual.
 
\section{Conclusions}
In conclusion, we demonstrated in this paper that the standard approach to describe the magnetism in solids with strongly correlated electrons, proceeding from the isolated ions in their ground state, and adding electron hopping between ions perturbativly, may break down in certain situations, especially for $4d$ and $5d$ systems. In particular, due to large $d-d$ hopping orbital-selective covalent bonds, or singlet molecular orbitals between transition metal ions may form in this case. This would lead to a strong suppression of the effective magnetic moment, and, for fractional occupation of $d-$shells, can strongly reduce or completely suppress the well-known double exchange mechanism of ferromagnetism. Several examples of real material considered in our paper indeed demonstrate that this mechanism is very efficient in suppressing the double exchange in a number of $4d$ and $5d$ compounds, and can even operate in some $3d$ systems. Spin-orbit coupling, especially relevant for $4d$ and $5d$ compounds, can also play an important role and in many cases it works together with electron hopping to suppress magnetic state, although in some situations it can also act in opposite direction. Our results show yet one more nontrivial effect in the rich physics of systems with orbital degree of freedom, especially those close to Mott transition.

\section{Acknowledgements} 
This work was supported by the Russian Foundation of Basic Research via Grant No.
16-32-60070, Civil Research and Development Foundation via program
FSCX-14-61025-0, FASO (theme ``Electron'' No. 01201463326), Russian ministry of education and science via act 11 contract 02.A03.21.0006, by K\"oln University via the German Excellence Initiative, and by the Deutsche Forschungsgemeinschaft through SFB 1238.

\bibliography{../../../library}
\nocite{MommaK.Izumi2011}
\nocite{Hirsch1986}
\nocite{Perdew1996}
\nocite{Blaha2001}

\end{document}